\documentclass{JINST}

\preprint{LAL 10-44}

\title{Per Mill Level Control of the Circular Polarisation of
the Laser Beam for a Fabry-Perot Cavity Polarimeter at HERA}

\author{V.~Brisson, R. Chiche, M.~Jacquet\thanks{Corresponding author.}, 
C.~Pascaud, V.~Soskov, 
 Z.~Zhang, F.~Zomer\\
\llap{}Laboratoire de l'Acc\'el\'erateur Lin\'eaire, Univ.\ Paris-Sud et 
IN2P3/CNRS,\\
  Orsay, France\\
  E-mail: \email{mjacquet@lal.in2p3.fr}}

\author{M.~Beckingham\thanks{Now at Albert-Ludwigs-University Freiburg, Freiburg, Germany.}, N.~Coppola\thanks{Now at European XFEL GmbH, Hamburg, Germany.}\\
\llap{}DESY\\
  Hamburg, Germany}

\abstract{A precise and fast Fabry-Perot cavity polarimeter, installed
in the HERA tunnel in the summer of 2003, was used to measure
the longitudinal polarisation of the lepton beam.
A complete theoretical model has been developed in order to control 
at the per mill level the degree of circular polarisation of the laser beam.
The transport of this quantity
within the whole optical setup has also been performed and controlled 
at the same level of precision.
This is the first time that such a precision is achieved in the difficult, 
hostile and noisy environment of a particle collider.}

\keywords{Particle collider, Polarimeter, Ellipsometer, Light polarisation, Quarter Wave Plate}

\begin{document}

\section{Introduction}
In a companion article \cite{polca-paper1} we describe 
a Compton polarimeter installed at HERA using a Fabry-Perot resonator 
to enhance the laser beam power. A relative statistical precision of 
$2\%$ per bunch and per minute was achieved on the longitudinal 
polarisation $P_z$ measurement of the electron beam with an estimated 
relative systematic uncertainty of about $1\%$. 
One of the systematic error sources is related to the determination of 
the degree of circular polarisation $S_3$ of the laser beam at 
the electron-laser interaction point (IP). Since only the product 
$S_3P_z$ is determined in Compton polarimetry (see Eq.(5) 
in~\cite{polca-paper1}), we have to measure precisely
$S_3$ in order to achieve the same level of precision
for $P_z$.
The uncertainty on the $S_3$ determination comes mainly from two sources: 
(1) the measurement of $S_3$ itself performed with an ellipsometer 
usually located close to, but outside, the electron beam pipe
and, (2) the transport of the measured $S_3$ value through optical elements 
and vacuum window up to the electron-laser IP. The purpose of the present
article is to describe the experimental setup and methods that we have used 
to reach a few per mill level of systematic uncertainty on $S_3$.

The key component of the experimental setup~\cite{polca-paper1}
is an ellipsometer similar to those of the SLAC~\cite{slac1,slac2} and 
Jlab~\cite{these-falletto-1999,fabry-perot-2001} Compton polarimeters. 
It is composed of a quarter wave plate (QWP), 
a linear polariser and various photo-detectors.
Since an optical model is needed to reconstruct $S_3$ from the photometric 
measurements performed after the polariser, the model
accuracy has to be controlled below the per mill level. 
The QWP is a crucial component of the ellipsometer. It is usually 
anti-reflection coated with double layers and thus taken as  
a simple delay plate in basic optical models~\cite{jones-I-1941}.
However, the reflectance of such coated plates is typically
of the order of $0.5\%$, thus limiting the model accuracy to the same level.
In order to decrease the model uncertainty, we 
followed the work of~\cite{poirson-et-all-gaussian-beam} by choosing 
an uncoated quartz QWP of high optical quality.
In doing so we have to account for multiple reflections
inside the anisotropic uniaxial QWP, to model the plate defects and 
the experimental misalignments, and to perform a fine
calibration of the plate thickness. 
The implementation of an uncoated QWP in the ellipsometer together with
a thorough investigation of theoretical models and detector effects 
within an accelerator environment has never been
reported previously and is one of the main topics of this article.

In order to control the transport of accurately measured $S_3$
up to the IP, we follow the method developed for 
the polarimeter at SLAC~\cite{slac} by modeling the optical elements located 
between the IP and the ellipsometer, and
use optical theorems demonstrated by Jones~\cite{jones-II-1941}. 
In an accelerator this transport is an important issue
since optical elements are always present between the IP and the ellipsometer.
All birefringence biases from these elements are therefore studied 
and/or modeled to conserve the required precision on $S_3$ at the IP.
We also perform a polarisation transport study from the IP up to the laser
head by using the optical theorems in~\cite{aspect-1993}. 
The implementation of these methods is the second topic of this article. 
To our knowledge they have not been applied in the context of accelerators to 
the level of accuracy presented here.
 
The ellipsometer and the characterisation of its optical components
are described in Sect.~\ref{chapter-ellipso}.
In Sect.~\ref{elli-systematics}, the determination of $S_3$
with the ellipsometer and the transport
from the ellipsometer to the electron-laser IP are described. 
Finally, in Sect.~\ref{aller-retour}, the characterisation of 
the entrance optical line is performed in order to study
the coherence of $S_3$ along the whole optical system
and in particular the coherence of $S_3$ at the entrance and at the exit of
the Fabry-Perot cavity.

\section{Ellipsometer characterisation}\label{chapter-ellipso}

In this section, the general optical setup and the characterisation of the
ellipsometer are described.
The principle of an ellipsometer is to send a light beam, of any unknown polarisation, through
a QWP. By rotating the plate, the polarisation state of the light is modified
and the state at the exit of the plate depends on the state at the entrance. 
A polariser (Wollaston prism) placed  behind the plate spatially separates the beam into
two orthogonal linearly polarised states. The analysis of the intensities
of these two beams in photo-detectors, for various azimuthal angles of
the QWP, allows the deduction of the polarisation of the incident beam.

\subsection{General optical setup and ellipsometer components}\label{exp-setup}

\subsubsection{General optical setup}

\begin{figure}[htbp]
\begin{center}
\includegraphics[width=0.65\textwidth]{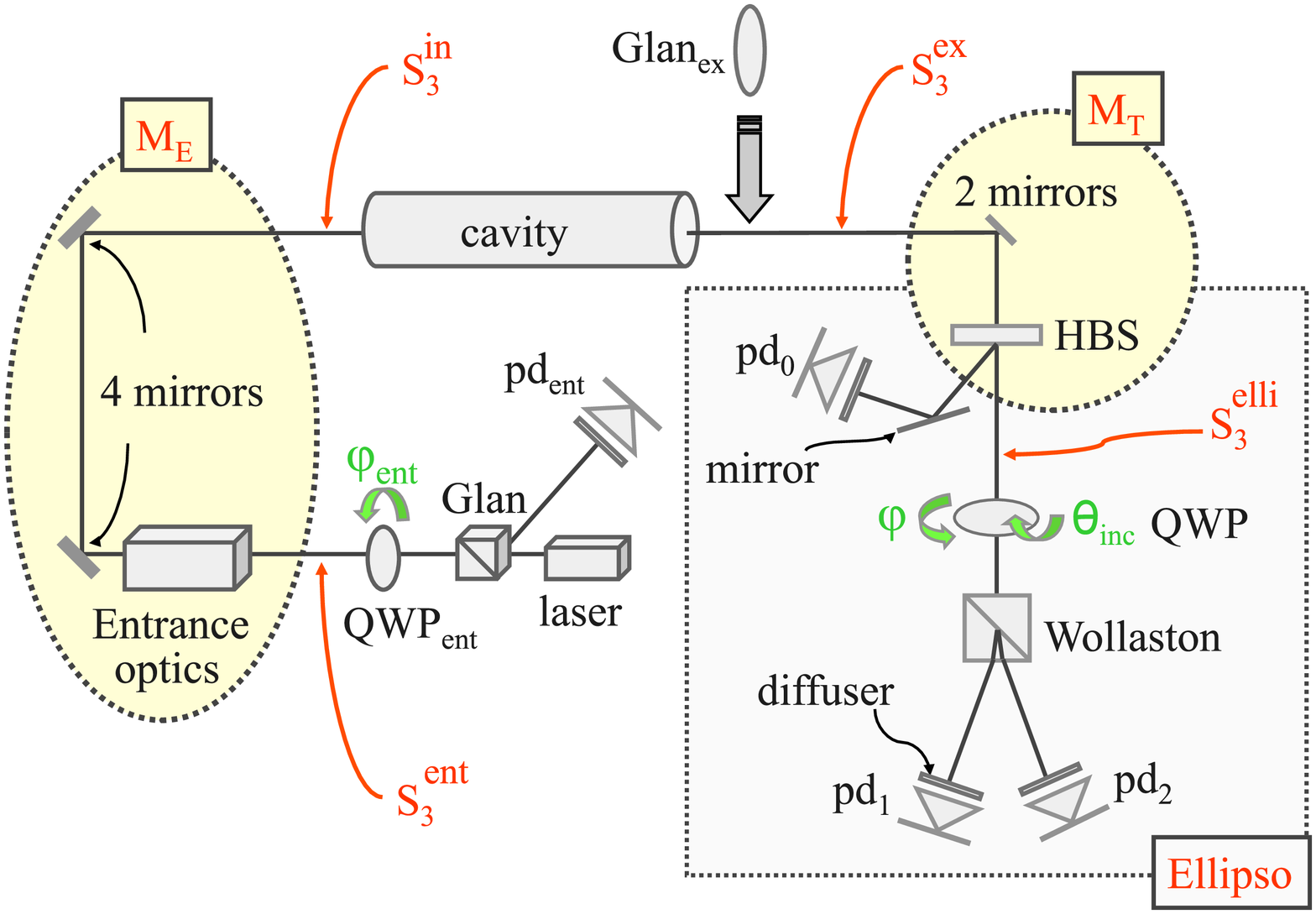}
\end{center}
\caption{
A schematic view of the Fabry-Perot cavity optical system installed 
in the HERA tunnel. The box ``entrance optics'' is composed of a glass plate 
and two lenses. The location of four determination points of $S_3$ is 
indicated by $S_3^{\rm ent}$, $S_3^{\rm in}$, $S_3^{\rm ex}$ and 
$S_3^{\rm elli}$. The prism $\mathrm{Glan_{ex}}$ located
on the top of the figure is inserted in the optical line only for 
a dedicated study described in 
Sect. 3.2.2. 
$M_E$ and $M_T$ are two transfer matrices discussed in
Sects. 3 and 4
.}
\label{shema-general}
\end{figure}

A schematic overview of the Fabry-Perot cavity optical setup is presented in
Fig.~\ref{shema-general}.
A Nd:YAG laser beam of $1064\,{\rm nm}$ wavelength passes first through a 
Glan-Thomson prism in order to provide a purely linearly polarised state
and then through an entrance QWP, noted $\mathrm {QWP_{ent}}$. 
This is mounted at the center of a motorised rotating stage to adjust the 
azimuthal angle $\phi_{\rm ent}$
and thus to provide an elliptical polarised state. The Glan allows also
the beam to pass and to go back after its reflection by the cavity 
entrance mirror.
The reflected beam is analysed in a photodiode, $\mathrm {pd_{ent}}$,
first to determine the azimuthal angles of the plate $\mathrm {QWP_{ent}}$
for which the light is circularly polarised and secondly to conduct 
a study on the entrance optical line (Sect.~\ref{me-determination}).
The beam then passes through the entrance optics which is 
composed of a glass plate and two lenses.
The glass plate is used to pick up a fraction of the beam for 
the locking procedure exploiting the ``Pound-Drever''
technique~\cite{feedback-black-2000},
and the lenses are used to match the laser beam to the cavity fundamental mode.
The beam is then precisely aligned with four mirrors (of which two are motorised) before 
entering a two meter long cavity.

At the exit of the cavity, the beam is guided with two mirrors to enter the ellipsometer,
labeled ``Ellipso'', which is mainly composed of a QWP followed by a Wollaston prism.
The beam passes first through a holographic beam 
sampler (HBS) in order to extract a small fraction of the entrance power
(about $1\%$ of the incident beam).
This fraction of the beam is measured in photodiode ${\rm pd}_0$ and
is used as a reference intensity to compensate the effects due to possible
laser power variations.
The main beam emerging from the HBS enters the QWP.
The QWP is mounted at the center of a motorised rotating stage to adjust the 
azimuthal angle $\phi$. 
Two screws allow the alignment between the plate and the rotating
stage to be adjusted. The rotating stage itself is mounted on a two-axis 
horizontal stage, a vertical translation stage and an angle-tilting stage 
to position the beam impact point at the mount center of the rotating stage
and to adjust the incident angle $\theta_{\rm inc}$ between the
laser beam and the plate normal direction. 
The linear and tilted stages are manually controlled with micrometric screws.
The Wollaston prism separates the laser beam spatially into two linearly and 
orthogonally polarised components, and
the two transmitted beams from the Wollaston are detected in photodiodes ${\rm pd}_1$ 
and ${\rm pd}_2$.
Diffusers are placed in front of each photodiode in order 
to adjust the power entering the photodiodes.

The ellipsometer is used to measure the circular polarisation 
$S_3$ of the laser beam of any unknown polarisation
by varying the azimuthal angle of the plate and analysing
the intensities emerging from the prism.

\subsubsection{Ellipsometer components}\label{elli-component}

{\it The HBS:} 
The beam sampler is a hologram in engraved relief on a transparent substrate (silica) 
inducing forward diffraction.
The birefringence of the HBS alone has been measured before its installation 
in the cavity system and the result is compatible with 
zero~\cite{linz-measure}.

\vspace{0.3cm}

\noindent {\it The Wollaston prism:} 
The most important characteristic of the prism is its extinction rate,
which is less than a few $10^{-5}$ as given by the manufacturer and 
has been confirmed experimentally.

\vspace{0.3cm}

\noindent {\it The quarter wave plates:}
As will be described in Sect.~\ref{qwpdetermination},
two QWPs ${\rm pl}_1$ and ${\rm pl}_2$
with different nominal thicknesses of
$e_{\rm nom}^{(1)}=91.2\,\mu{\rm m}$ and $e_{\rm nom}^{(2)}= 639.9\,\mu{\rm m}$
are used in the ellipsometer 
for its characterisation in order
to increase the constraints of the system.
Each QWP is a parallel plate of high quality quartz manufactured 
especially for the ellipsometer characterisation purpose, and
has a delay tolerance of $1/300$ and a thickness tolerance of a few 
micrometers with a parallelism between the two faces of the order of 
$10$ seconds of arc.
The optical axis is contained in the plane of the plate.

\vspace{0.3cm}

\begin{figure}[htbp]
\begin{center}
\includegraphics[width=0.495\textwidth]{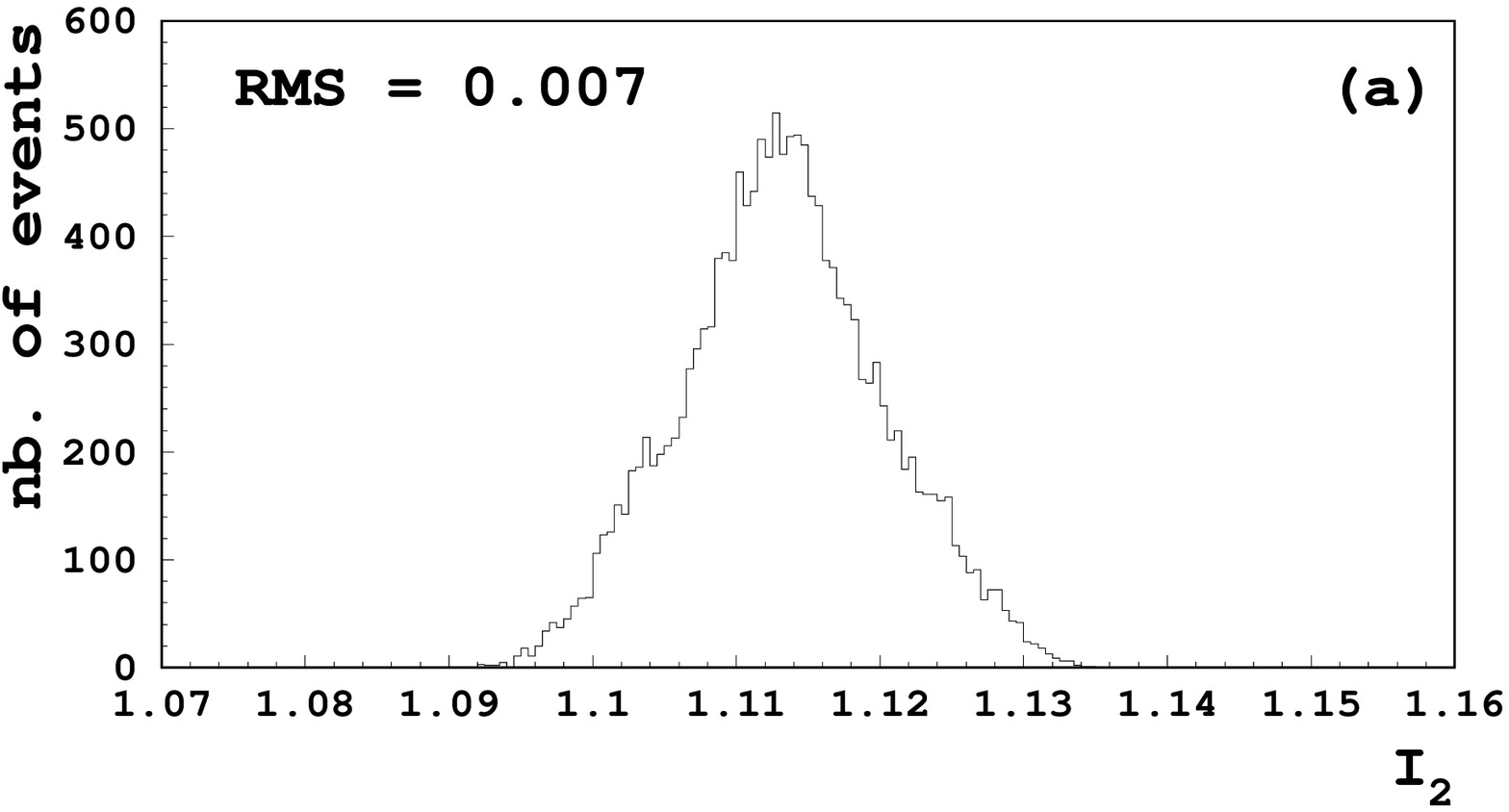}
\includegraphics[width=0.495\textwidth]{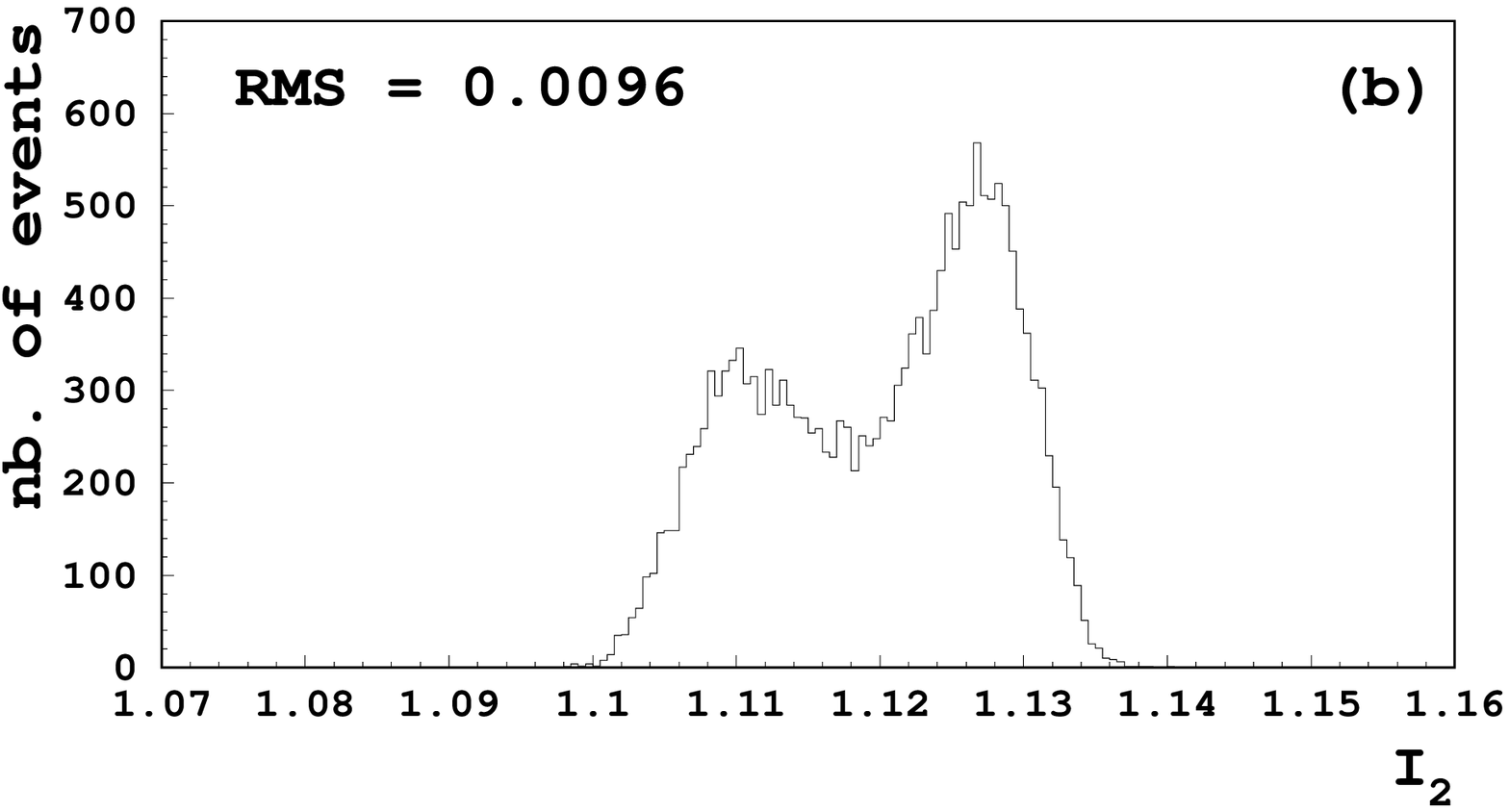}
\end{center}
\caption{ Histograms of $I_2$ as defined in 
Eq.(2.1)
for two data sets (a) and (b) taken during fifteen minutes each
at different moments.}
\label{data-histo-desy}
\end{figure}

\begin{figure}[htbp]
\begin{center}
\includegraphics[width=0.495\textwidth]{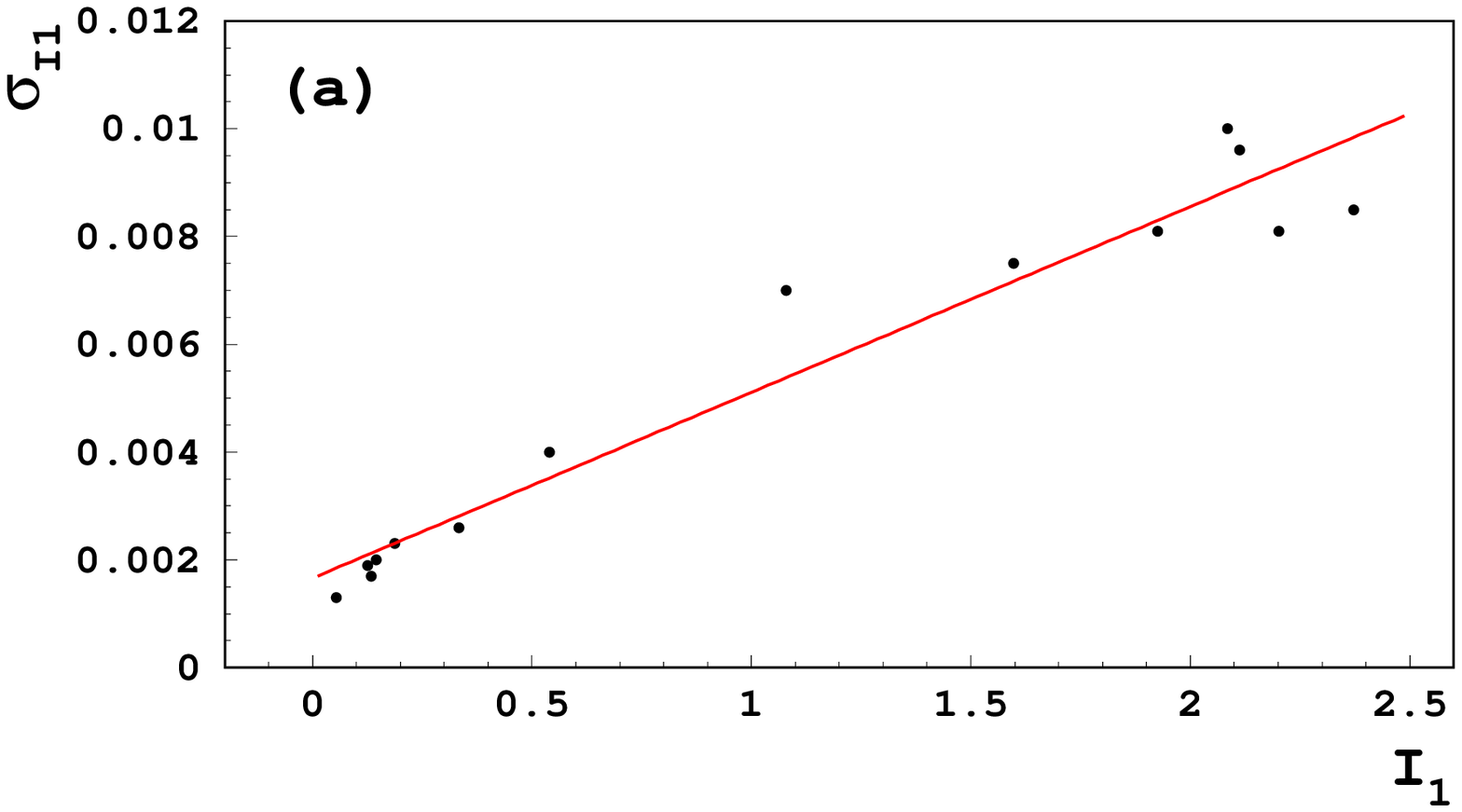}
\includegraphics[width=0.495\textwidth]{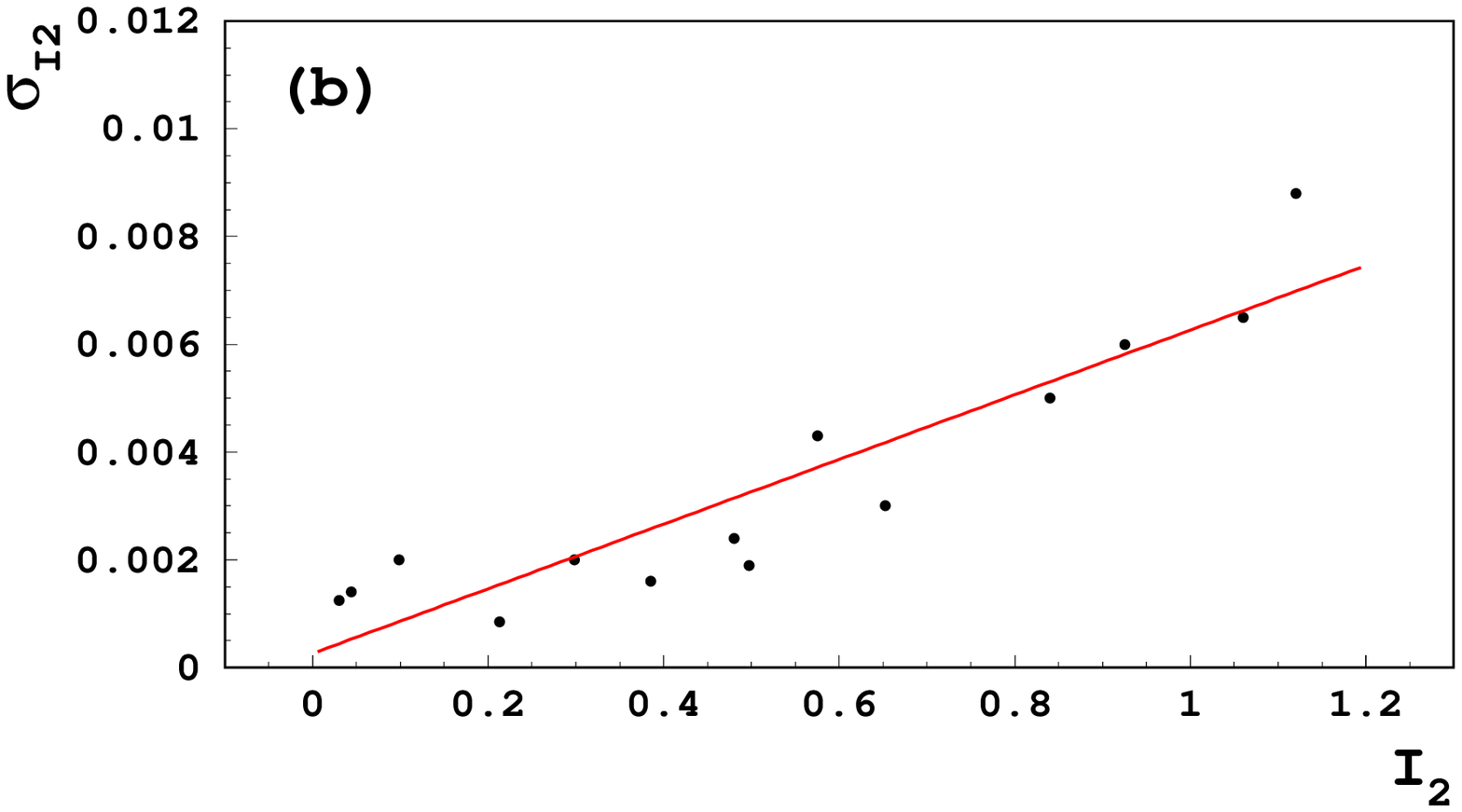}
\end{center}
\caption{Errors $\sigma_{I_1}$ (a) and $\sigma_{I_2}$ (b)
as a function of $I_1$ and $I_2$, respectively.
The lines correspond to straight line fits to the measured errors according to 
Eq.(2.2)
.}
\label{data-error-desy}
\end{figure}

\noindent {\it The detection system:}
The detection system consists of three photodiodes
${\rm pd}_0$, ${\rm pd}_1$ and ${\rm pd}_2$
made of a gallium arsenide and indium alloy (InGaAs).
Each photodiode and their electronics are thermalised with a Peltier module 
to prevent temperature variations which might be of a few degrees in the HERA
environment and could induce
a variation in the photodiode readout larger than the required precision.
Using Peltier modules, a stability level of tenth of a degree is achieved.
The photodiodes are read out with $12$-bit analog-to-digital converters at 
a maximum sample rate of $2\,{\rm MHz}$. For each measurement, the mean value 
over ten thousand signals $I_{\rm pd}$ is recorded after
subtraction of a pedestal ${\rm Ped}$ for each photodiode.
The effect of laser power variations is taken into account by normalising
the measurement to that of the reference photodiode ${\rm pd}_0$. 
Explicitly, photodiode intensities $I_1$ and $I_2$ used in the following 
can be written as: 
\begin{equation}
\label{photodiodes-sig}
I_{1,2} = \frac{ \left< I_{{\rm pd}_{1,2}} \right>_{10000} - {\rm Ped}_{1,2}}
               { \left< I_{{\rm pd}_0} \right>_{10000} - {\rm Ped}_0 }\,.
\end{equation}
A study of photodiodes in an optical laboratory has shown that in such a clean 
environment, the temperature regulation, the subtraction of photodiode 
pedestals and the laser power variation effect can be controlled 
such that $I_1$ and $I_2$ are known at the per mill level.
The HERA accelerator environment is more noisy;
effects such as larger temperature variations and
the presence of quadrupoles and dipoles, the synchrotron radiation,
accelerating cavities and long cables
affect the precision of the photodiode measurements.
This is illustrated in Figs.~\ref{data-histo-desy}(a) and (b)
where the distributions of $I_2$ from two data samples recorded
in the HERA tunnel at different moments are shown. 
Each entry in the histograms is a measurement of 
$I_2$ as defined in Eq.(\ref{photodiodes-sig}) and
the duration of data taking for each sample was approximately fifteen minutes. 
Fig.~\ref{data-histo-desy}(a) represents a well clustered
distribution whereas Fig.~\ref{data-histo-desy}(b) shows
two populations.
To take into account this kind of drift,
the corresponding error $ \mathrm{ \sigma_{I_{1,2}}}$
is defined as the root mean square (RMS) value of the histogram,
whether it has a Gaussian shape or not.
In this way, the unknown long term effects
such as those illustrated in Fig.~\ref{data-histo-desy}(b) are included 
in the measurement errors.
The errors are then parameterised as a function of the intensities as: 
\begin{equation} \label{eq:error}
\sigma_{I_{1,2}= A_{1,2}\ I_{1,2} + B_{1,2}}\,.
\end{equation}
Fig.~\ref{data-error-desy} shows the errors $\sigma_{I_1}$ (a) and 
$\sigma_{I_2}$ (b) as a function of $I_1$ and $I_2$ respectively and 
the corres-
ponding parameterisation of Eq.(\ref{eq:error}),
for data sets recorded at various azimuthal angles of the ellipsometer QWP 
in order to cover the entire range of intensity values. 
These errors may depend on the duration of acquisition time which
will be varied for systematics studies (Sect.~\ref{docp-measurements}).

The accurate measurement of the degree of circular polarisation $S_3$
of the laser beam requires a precise control and characterisation of this
ellipsometer.
Therefore before determining $S_3$, the ellipsometer will be first used as 
a calibration system to characterise precisely some of its optical 
components. 
For this, a complete simulation model has been developed.

\subsection{Model of the ellipsometer and $\chi^2$ function}\label{simul-model}

\subsubsection{Model of the ellipsometer}

The polarisation state ${\bf E}_{\rm elli}$ at the entrance of 
the ellipsometer and the associated degree of circular polarisation 
$S_3^{\rm elli}$ can be written in the most general form as functions of 
two angles $\xi_{\rm elli}$ and $\phi_{\rm elli}$ 
as~\cite{bouquin_serge_huard}:
\begin{equation}\label{eq:docp}
 {\bf E}_{\rm elli} = \left( \begin{array}{l} 
 \cos\xi_{\rm elli} \\
 \sin\xi_{\rm elli}\ e^{i \phi_{\rm elli}} \end{array} \right)\,,
\hspace{0,5cm}
S_3^{\rm elli} = 2 \cos \xi_{\rm elli}\sin \xi_{\rm elli}\sin \phi_{\rm elli}\,.
\end{equation}
The expression of the two transmitted fields ${\bf E}_1$ and ${\bf E}_2$ 
emerging from the Wollaston prism and the associated theoretical intensities 
$T_1$ and $T_2$ can then be written as:
\begin{equation}\label{eq:i12trans}
{\bf E}_{1(2)} = {\mathcal M}^{\rm elli}_{1(2)}
{\bf E}_{\rm elli}\,,\hspace{0,5cm}
T_{1(2)} = \left| {\bf E}_{1(2)} \right|^2\,,
\end{equation}
where ${\mathcal M}^{\rm elli}_1$ and ${\mathcal M}^{\rm elli}_2$,
standing for the corresponding Jones transmission matrices~\cite{jones-V-1947},
are computed from a theoretical model as follows:

\vspace{0.3cm}

\noindent {\it The Wollaston cube:}
Two small parameters $\epsilon_x$ and $\epsilon_y$ are introduced to 
take into account in the Wollaston Jones matrix 
a possible small birefringence along its two optical 
axes~\cite{bouquin-azzam-et-bashara}.

\vspace{0.3cm}

\noindent{\it The quarter wave plate:} 
The model used for the QWP takes into account 
the multiple reflections of the wave inside a quartz plate of indices 
$n_o$ and $n_e$, of thickness $e_{\rm QWP}$, and whose optical axis has 
an angle $\phi_{\rm oa}$ with respect to the laboratory frame.
This angle has two components: $\phi_{\rm oa} = \phi + \phi_0$,
where $\phi$ is the azimuthal angle of the QWP motorised rotating stage and
$\phi_0$ is an azimuthal reference angle reflecting the fact that 
the orientation of the optical axis in the plane of the plate 
is not a priori at position $\phi=0$.
The simulation also models the passage of a light wave through the plate 
at a non zero incident angle $\theta_{\rm inc}$, which is equal to
the number of tilting stage screw turns ($X_{\rm turn}$) times 
the tilt angle value of one screw turn ($\theta_{\rm turn}$).
The light beam is treated as a plane wave since, at small incident angles
({\it i.e.} less than $0.1\,{\rm rad}$ for the work described here), 
the comparison between a plane wave treatment and a Gaussian wave treatment
shows that the Gaussian character of the wave can be 
neglected~\cite{article-fabian-gaussian-beam}.
Also, the contribution of the optical activity of 
the crystal~\cite{article-fabian-optical-activity} 
as well as the surface roughness~\cite{article-fabian-roughness}
has been studied and 
found to be negligible (a relative contribution of less than $10^{-6}$). 

\vspace{0.3cm}

\noindent{\it Misalignments:}
A realistic description of the optical system must also 
take into account the following experimental misalignments:
\begin{itemize}
\item A misalignment of the QWP plane with respect to the Wollaston
prism axes, which is mo-
deled by a small tilt angle $\delta_W$ of the normal to
the QWP.
\item A misalignment due to the fact that the laser beam may not enter 
the QWP exactly at the plate center of the mount: in that case, 
because the two plate surfaces may not be perfectly parallel, 
the thickness crossed by the light can vary during an azimuthal rotation of 
the plate. To take into account this effect, the plate is modeled by 
a variable thickness as:
$e = e_{\rm QWP}\ [ 1 + (\delta_{c}/e_{\rm QWP})$ 
$\cos(\phi-\phi_{c}) ]$ where $e_{\rm QWP}$ is the thickness of the plate 
in case of a perfect alignment situation, $\delta_{c}$ represents 
a small shift between the laser beam impact point and the plate center of 
the mount, and $\phi_{c}$ is an arbitrary reference azimuthal angle, 
as the position $\phi=0$ of the QWP may not be 
the position which maximises the plate thickness $e$.
One new parameter $\delta_{c}$ is introduced each 
time the incident angle $\theta_{\rm inc}$ changes.
\end{itemize}

According to this model of ellipsometer components and misalignments, 
for a given ellipsometer QWP plate ${\rm pl}_k$ of thickness $e^k_{\rm QWP}$,
a given incident angle $\theta_{\rm inc}^j$, azimuthal angle
$\phi^i$ of the plate, and polarisation state 
${\bf E}_{\rm elli}^{\ell}$,
the two theoretical intensities $\mathrm{T_{1(2)}}$ of Eq.(\ref{eq:i12trans})
can be written as functions of ellipsometer parameters as:
\begin{equation}\label{elli-param}
\begin{array}{ll}
T_{1(2)}^{ijk\ell} \equiv f_{1(2)}\ 
(n_o,& 
  n_e,\, \epsilon_x,\, \epsilon_y,\, e^k_{\rm QWP},\, \theta_{\rm turn} \times X_{\rm turn}^j, \\
 & \phi^k_0+\phi^i,\,  \xi^{\ell}_{\rm elli},\, \phi^{\ell}_{\rm elli},\,
\delta^k_W, \delta^j_c,\, \phi^k_c )\,.
\end{array}
\end{equation}
All the details of the model described here can be found in~\cite{habil-marie} 
where the modeling and the calculations relative to the Wollaston cube, 
the QWP and the optical misalignments are explicitly given. 
In particular, the calculation of the Jones transmission matrix of 
a quartz plate at a non zero incident angle, with internal reflections 
being taken into account, is explicitly performed in the appendix 
of~\cite{habil-marie}.
Simulations show that multiple reflections, plate defects or 
optical misalignments contribute at the few percent level to 
the transmitted intensities $T_{1(2)}$.
Achieving the per mill level for $S_3$ measurement is therefore
only possible using this complete model description.

\subsubsection{$\chi^2$ function}

To characterise the ellipsometer and distinguish
effects due to optical misalignments from those due to plate defects or those
due to the light polarisation state, the principle
is to record ellipsometer experimental data $I_{1(2)}$ 
(as defined in Eq.(\ref{photodiodes-sig}))
and to minimise the following $\chi^2$ function:
\begin{equation}\label{eq:chi2pola}
\begin{array}{ll}
{\displaystyle \chi^2 = } & 
{\displaystyle \sum_{k=1,N_{\rm pl}}\,\sum_{\ell=1,P_k}\,
\sum_{j=1,\Theta_k}\,\sum_{i=1,N_{\phi}}} \\ &
\left[ \left( {\displaystyle \frac{R_1^{j \ell k} T_{1}^{ij \ell k}-
 I_{1}^{ij \ell k}}
{\sigma_{1}^{ij \ell k}}} \right)^2  +
\left( {\displaystyle \frac{R_2^{j \ell k} T_{2}^{ij \ell k}-I_{2}^{ij \ell k}}
{\sigma_{2}^{ij \ell k}}} \right)^2 \right]\,,
\end{array}
\end{equation}
where $N_{\rm pl}$ refers to the number of the QWP used in the data taking, 
$P_k$ to the number of polarisation states,
$\Theta_k$ to the number of incident angles and
$N_{\phi}$ to the number of azimuthal angles of the QWP.
$T_{1(2)}^{ij \ell k}$ ($I_{1(2)}^{ij \ell k}$) is the theoretical 
(experimental) photodiode intensity calculated (measured) 
when the plate ${\rm pl}_k$ is inserted in the ellipsometer, 
at the $\ell^{\rm th}$ polarisation state ${\bf E}_{\rm elli}$,
the $j^{\rm th}$ value of $\theta_{\rm inc}$ 
and the $i^{\rm th}$ value of $\phi$.
$\sigma_{1(2)}^{ij \ell k}$ is the uncertainty of $I_{1(2)}^{ij \ell k}$
in Eq.(\ref{eq:error}).
$R^{j\ell k}_{1(2)}$ are normalisation factors which are easily determined
by solving $\partial \chi^2 / \partial R =0$.
The minimisation of the $\chi^2$ (Eq.(\ref{eq:chi2pola})) leads to
parameter values of Eq.(\ref{elli-param}),
excepted for $X_{\rm turn}^j$,  $\phi^i$ and $n_e$.
$X_{\rm turn}^j$ and $\phi^i$ take some known values of the tilting and 
rotating stages, respectively.
The extraordinary index $n_e$ is derived from previous 
studies~\cite{quartz-indice-carvallo,quartz-indice-ghosh}, where
the quartz indices were measured at a few $10^{-5}$ level,
because our constraints are not sufficient to determine $n_o$ and $n_e$ 
at the same time, and so the quantity that is determined in practice is 
the birefringence $n_e-n_o$. 
The determination of $n_o$ by the ellipsometer 
does not aim for the same level of precision as was obtained 
in~\cite{quartz-indice-carvallo,quartz-indice-ghosh} but
provides a good test of the validity of the model.

\subsection{Ellipsometer parameter determination}\label{qwpdetermination}

The determination of ellipsometer parameters (Eq.(\ref{elli-param})) has been 
performed twice, independently 
in two different environments corresponding to 
the optical laboratory and the HERA tunnel, 
with two different data taking procedures.

The first data sample recorded in the optical laboratory
is devoted to determine all ellipsometer characteristics and in particular 
the thickness of the QWP and the index $n_o$.
The determination of both $e_{\rm QWP}$ and $n_o$
requires a long and meticulous data taking procedure
since it turns out that the solution of the $\chi^2$ minimisation is not
unique: several combinations ($e_{\rm QWP},n_o$) can minimise the $\chi^2$.
In order to resolve such an ambiguity, 
two uncoated plates ${\rm pl}_1$ and ${\rm pl}_2$ with different thicknesses
are used by inserting one after the other in the ellipsometer.
In addition, for each plate, several data sets are taken for different 
incident angles. 

The second data sample was recorded after the installation of the setup 
in the HERA tunnel in order to characterise again the system, 
since all the optical components were dismounted to be
transported from the optical laboratory to the tunnel.
In the tunnel, data taking conditions were much more difficult than 
in the optical laboratory, and in particular, the tunnel accesses were 
limited to a few hours per month. 
The corresponding data taking procedure thus has to be simpler.
To avoid ambiguous solutions on the thickness of plate ${\rm pl}_1$, 
its fit range is restricted to around the expected value
obtained from the laboratory calibration.

\subsubsection{The data calibration samples}\label{datataking}

The first calibration of the ellipsometer was performed in the clean 
optical laboratory, where the room temperature was regulated at $25^\circ$. 
For a given laser beam polarisation state,
ellipsometer measurements were recorded for each plate ${\rm pl}_1$ and 
${\rm pl}_2$ at various incident angles $\theta_{\rm inc}$ between
the laser beam and the plate normal direction, and, 
for each $\theta_{\rm inc}$, the QWP was turned azimuthally of an angle 
$\phi$ in step of $1^\circ$ from $0^\circ$ to $360^\circ$. 
A Monte Carlo study shows that to provide enough constraints to minimise 
the $\chi^2$ and determine all the ellipsometer parameters of 
Eq.(\ref{elli-param}), data have to be recorded at 
four (two) different values of $\theta_{\rm inc}$
for the QWP ${\rm pl}_1$ (${\rm pl}_2$),
and it is sufficient to have only one polarisation state of a given 
$(\xi_{\rm elli},\phi_{\rm elli})$.
Each time the incident angle $\theta_{\rm inc}$ was changed, 
a long procedure was applied to displace manually the plate transversally
with the linear stage micrometric screws
in order to recover precisely the matching of the plate mechanical center with 
the laser impact point.
The six samples recorded at different values of $\theta_{\rm inc}$ 
containing $360$ entries each are called $D^{\rm lab}_d$ ($d=1,\cdots,6$) and,
when introduced in the $\chi^2$ function, values of subscripts in 
Eq.(\ref{eq:chi2pola}) are $N_{\rm pl}=2$, $P_1=1$, 
$P_2=1$, $\Theta_1=4$, $\Theta_2=2$ and $N_{\phi}=360$.

The second data set was recorded in the HERA tunnel where
the room temperature was re-
gulated at $\mathrm{ 35^o}$.
In order to control the data taking procedure from outside tunnel to be 
independent of the short duration of tunnel access,
only one plate (${\rm pl}_1$) was used in the ellipsometer and 
the incident angle $\theta_{\rm inc}$ between the light beam and this QWP 
remained fixed and equal to zero. 
In this configuration, the $\chi^2$ minimisation was performed by
using three recorded data samples, each with
a different azimuthal angle of the entrance plate ${\rm QWP}_{\rm ent}$.
In this way, the light entering the ellipsometer has three different 
polarisation states.
For each of these three ${\rm QWP}_{\rm ent}$ azimuthal positions,
the ellipsometer QWP was also turned azimuthally through an angle $\phi$ 
in step of $1^\circ$ from $0^\circ$ to $360^\circ$.
These three data samples are called $D^{\rm HERA}_d$ ($d=1,\cdots,3$)
and the corresponding values for the superscripts in Eq.(\ref{eq:chi2pola}) are:
$N_{\rm pl}=1$, $P_1=3$, $\Theta_1=1$ and $N_{\phi}=360$.

\subsubsection{Correlation between $e_{\rm QWP}$ and $S_3$}\label{correl}

\begin{figure}[htbp]
\begin{center}
\includegraphics[width=0.55\textwidth]{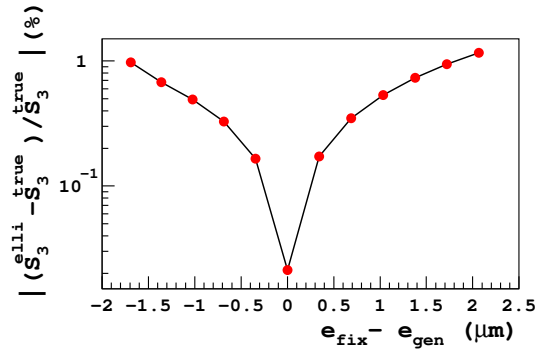}
\end{center}
\caption{Relative error on $S_3^{\rm elli}$ as a function of 
the uncertainty on the ellipsometer QWP thickness.}
\label{de-dp3}
\end{figure}

Among all ellipsometer parameters, the dominant source of systematic error on 
$S_3$ comes from the QWP thickness uncertainty. 
Thus, before giving the results of the minimisations using the two sets of 
data samples previously described,
it is interesting to show the correlation between the plate thickness 
$e_{\rm QWP}$ and $S_3$.
The effect on the $S_3$ determination
is estimated by simulating an ellipsometer data sample 
with a degree of circular polarisation $S_3^{\rm true}$
and a plate thickness $e_{\rm gen}$.
Using this sample, various minimisations of the 
$\chi^2$ are performed by letting only $S_3^{\rm elli}$ free 
({\it i.e.} only the two angles $\xi_{\rm elli}$ and $\phi_{\rm elli}$), 
by fixing the plate thickness to different
values $e_{\rm fix}$ slightly different from $e_{\rm gen}$, 
and by fixing all the other parameters to their generated values.
The quantity $| (S_3^{\rm true}-S_3^{\rm elli}) /S_3^{\rm true} |$
presented in Fig.~\ref{de-dp3} as a function of $e_{\rm fix} - e_{\rm gen}$
shows that a systematic error of one micometer on the plate thickness 
leads to a systematic error around $0.5\%$ on $S_3$.
A precise knowledge of the plate thickness 
inside the thickness tolerance of a few micrometers given by the manufacturer
has therefore to be reached to keep a systematic error at the per mill level 
on the measurement of $S_3$.
Achieving this precision is only possible using 
the complete model described previously.

\subsubsection{Results}\label{calibfit}

\begin{figure}[htbp]
\begin{center}
\includegraphics[width=\textwidth]{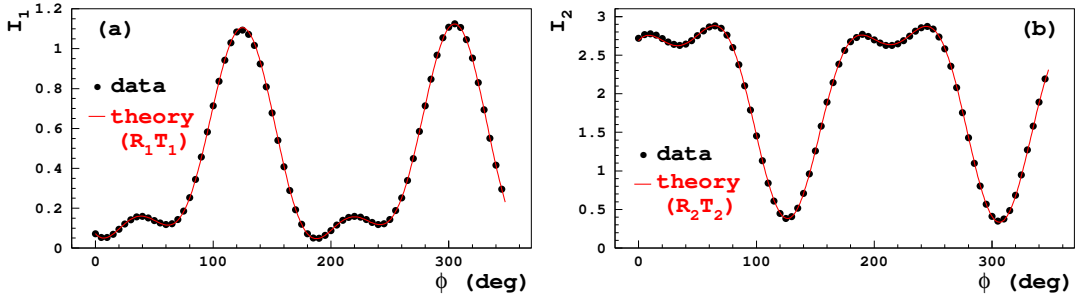}
\end{center}
\caption{Experimental intensities (black bullets, for clarity only a subsample
is shown) $I_1$ (a) and $I_2$ (b) 
compared with theoretical ones (curves) derived from the fit,
as a function of the azimuthal angle $\phi$ of the ellipsometer QWP, 
for the data file $D^{\rm HERA}_1$.}
\label{mc-data-pola-i-phi}
\end{figure}

The $\chi^2$ minimisation procedure was tested first by using simulated 
samples generated with Eq.(\ref{eq:i12trans}) following closely the 
experimental data. It was found that all fitted parameters were 
in agreement with the generated ones within a precision of a few per mill.
The minimisation is then performed independently with 
the two experimental data sets $D^{\rm lab}_d$ and $D^{\rm HERA}_d$ and leads to values of
$\chi^2$ per degree of freedom equal to 1.07 and 2.09 respectively. 
The excellent agreement between experimental intensities and 
theoretical ones based on the fit is illustrated by a typical example 
in Fig.~\ref{mc-data-pola-i-phi}, where
the quantities $I_{1,2}$ and $R_{1,2}T_{1,2}$ are presented for the sample 
$D^{\rm HERA}_1$ as a function of the azimuthal angle $\phi$ of 
the ellipsometer QWP. 

\begin{figure}[htbp]
\begin{center}
\includegraphics[width=0.54\textwidth]{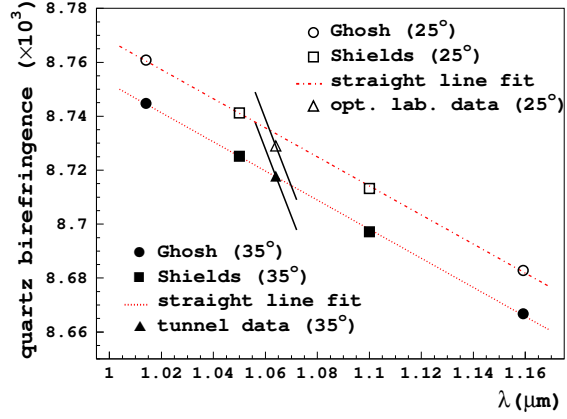}
\end{center}
\caption{Quartz birefringence values scaled to $25^\circ$ and $35^\circ$ from 
textbook values~\cite{quartz-indice-ghosh,quartz-indice-shields-and-ellis},
as a function of the wavelength. 
Dashed lines are straight line fits to four scaled textbook points.
Birefringence values determined from the fits to data $D^{\rm lab}$ and 
$D^{\rm HERA}$ are indicated by an open and a solid triangle,
respectively. The error bars are inclined for clarity.}
\label{biref2}
\end{figure}

All ellipsometer parameters determined from the two minimisations using
the data sets $D^{\rm lab}_d$ or $D^{\rm HERA}_d$ are found to be 
realistic and well defined.
Among them, one interesting quantity is the quartz birefringence value
$n_e-n_o$ which can be compared with textbook values previously determined.
In~\cite{quartz-indice-ghosh,quartz-indice-shields-and-ellis},
birefringence measurements were performed at a temperature of 
$18^\circ$ and $22^\circ$.
Our measurements in the optical laboratory and in the tunnel
were carried out at higher temperatures of $25^\circ$ and $35^\circ$, 
respectively. Based on the relation of optical index variation with 
temperature~\cite{quartz-indice-vs-temperature-toyoda-and-yabe},
quartz birefringence values of~\cite{quartz-indice-ghosh,quartz-indice-shields-and-ellis} 
are scaled to $T=25^\circ$ and at $T=35^\circ$
and are shown in Fig.~\ref{biref2}
together with the two birefringence values $n_e-n_o$, where the $n_o$ is 
obtained from the fits and the $n_e$ (and its uncertainty of $\sim 2\times 10^{-5}$) taken from 
Refs.~\cite{quartz-indice-carvallo,quartz-indice-ghosh}.
Our results agree at better than one per mill with the ones quoted 
in the references.

As for the birefringence, all other results determined in the optical 
laboratory are compared with those in the tunnel and
good agreement are found once the effects of 
temperature difference and optical alignment difference 
are taken into account~\cite{habil-marie}.


\section{Regular measurements of ${\bf S_3}$ and systematics studies}\label{elli-systematics}

\subsection{Measurements and systematic uncertainty from the ellipsometer}\label{docp-measurements}

During the data taking period of the cavity polarimeter, when 
the cavity was locked in resonant state and the azimuthal angle 
$\phi_{\rm ent}$ of the entrance QWP 
was such that the light is close to a fully right or left circularly
polarised state $S_3 = \pm 1$~\cite{polca-paper1},
values of $S_3^{\rm elli}$ were regularly determined.

Each value of $S_3^{\rm elli}$ is extracted from a data sample recorded with the
ellipsometer and
containing $180$ photodiode signals (as defined in Eq.(\ref{photodiodes-sig}))
corresponding to a $2^\circ$-step azimuthal turn of 
the ellipsometer QWP. The duration of data taking was about ten minutes 
for each sample.

To extract $S_3^{\rm elli}$ and its uncertainty 
$\sigma_{S_3^{\rm elli}}$, the $\chi^2$ function
defined in Eq.(\ref{eq:chi2pola}) is minimised.
The only fitted parameters are the polarisation state parameters 
$\xi_{\rm elli}$ and $\phi_{\rm elli}$ and all other
parameters are fixed to values previously determined by the characterisation 
of the tunnel optical system as described in 
Sect.~\ref{chapter-ellipso}.
Thanks to a heat regulation system, the temperature 
inside the isotherm house (see~\cite{polca-paper1}) was controlled within
$\pm 0.3^\circ C$ which ensured a perfect stability of the optical axis 
(defined by the positions of the cavity mirrors)
over time and therefore the stability of our ellipsometer calibration.

The uncertainty $\sigma_{S_3^{\rm elli}}$ obtained from the $\chi^2$ minimisation
is of the order of a few $10^{-4}$.
During one year (from June 2006 to June 2007), the
azimuthal positions of the plate ${\rm QWP}_{\rm ent}$
defining a left or right circularly polarised laser beam 
were changed only three times,
either after an hardware problem on the rotating mount controller of 
the plate, or after a dedicated check involving the plate or
the photodiode ${\rm pd}_{\rm ent}$.
The $S_3^{\rm elli}$ measurements were very stable over time since,
over the one-year period, they have shown a stability of 
a few per mill~\cite{habil-marie}.

Given the precise ellipsometer calibration procedure described in
Sect.~\ref{simul-model}, the only remaining source of systematic uncertainty 
on ${S_3^{\rm elli}}$ concerns the duration of the data taking.
Indeed, the duration of one measurement sample taken with the ellipsometer can 
have an effect on the photodiode intensity distributions because of long term 
fluctuation as shown in Fig.~\ref{data-histo-desy}(b).
This duration depends on the chosen interval between two consecutive 
azimuthal angles $\phi$ of the ellipsometer QWP.
To study this effect, the entrance plate ${\rm QWP}_{\rm ent}$ was fixed
to a given azimuthal position and several measurement samples 
were recorded using the ellipsometer for different durations
ranging from three to twenty minutes.
The extracted ${S_3^{\rm elli}}$ values from these samples are found to be
compatible within two per mill~\cite{habil-marie}.
To be conservative, an error of three per mill 
is quoted for the uncertainty due to the duration of data taking.

\subsection{Transport of $S_3$ from ellipsometer to cavity center} \label{docp-total-error}

The precise ${S_3}$ values obtained above concern the degree of circular
polarisation at the entrance of the ellipsometer.
What we are interested in is, however, the ${S_3}$ value 
at the electron-laser interaction point, {\it i.e.} at the center of the Fabry-Perot cavity.
A priori, these two values are the same, but because of the presence of
optical components between the two, a small difference could be induced.
The transport of ${S_3}$ and its uncertainty
are the subject of this section.

\subsubsection{Parasitic ellipticity from cavity center to ellipsometer entrance}\label{parasitic}

Between the center of the cavity and the entrance of the ellipsometer
(see Fig.~\ref{shema-general}), the exit mirror and the exit window of 
the Fabry-Perot cavity and the optical system $M_T$
could be a source of birefringence and 
may induce parasitic ellipticity, modifying the laser beam polarisation.
The birefringence of the substrate, the coating and the mounting system of the
exit mirror and the exit window has been estimated or measured from dedicated 
studies~\cite{fabhabil,habil-marie}. It was 
shown that the bias induced on ${S_3}$ from the center to 
the exit of the cavity is at the utmost of the order of $3 \times 10^{-5}$. 
The remaining dominant source of parasitic ellipticity is associated to 
the $45^\circ$ dielectric mirrors used to guide the light 
into the ellipsometer because of their different reflection coefficients for 
two electric field components: one in the incident plane and the other 
perpendicular to it~\cite{bouquin_serge_huard,bouquin-smith}.
To determine this ellipticity, the transfer matrix $M_T$ of the system has to be determined. 

\subsubsection{Exit transfer matrix $M_T$}\label{exit-matrice}

An optical theorem demonstrated by R.\ Clark Jones~\cite{jones-II-1941} states
that any optical system composed of any non-absorbing components may always 
be replaced by a partial polariser placed between two delay plates, 
with the addition of a rotator inserted at any position in the system.
Under the assumption that no power is absorbed in optical materials, 
the Jones matrices of a partial polariser, a delay plate and a rotator
${P_{{p_1}{p_2}}}$, ${ G_{\gamma}}$ and ${ R_{\theta}}$
can be written as
\cite{jones-I-1941}:
\begin{equation}\label{eq:matricepgr} 
 {P_{{p_1}{p_2}}=}\left(\begin{array}{cc}
{p_1} &  0 \\  0 & {p_2} \end{array}\right)\,,
\hspace{5mm}
{ G_{\gamma}=}\left(\begin{array}{cc}
{ e}^{i\gamma} &  0 \\  0 & {e}^{-i\gamma} \end{array}\right)\,,
\hspace{5mm}
{ R_{\theta}=}\left(\begin{array}{rr}
{ \cos \theta} & {  -\sin \theta} \\ {  \sin \theta} & {  \cos \theta}
\end{array}\right)\,.
\end{equation}
Following the Jones theorem~\cite{jones-II-1941}
and starting with a completely linear polarisation state
${ {\bf E}_G = }$ ${(\cos \phi_G ,\sin \phi_G )^T}$,
the theoretical final state ${ {\bf E}_{\rm th}}$ after passing through 
an optical system of transfer matrix ${M_T}$ 
can then be modeled as:
\begin{equation}\label{eq:descri-biref}
\begin{array}{ll}
& {\bf E}_{\rm th} \equiv (\cos \xi^{\rm th} ,\sin \xi^{\rm th}e^{{\it i}{\phi^{\rm th}}})^T
= M_T{\bf E}_G \\
{\rm with} \hspace{10mm}&
M_T=R_{\theta_1}\ G_{\gamma_1}\ R_{\theta_2}\ P_{{p_1}{p_2}}\ R_{\theta_3}\ G_{\gamma_2}\,.
\end{array}
\end{equation}

In order to apply this theorem to determining the matrix ${M_T}$ of 
the two mirrors and the HBS located at the exit of the Fabry-Perot cavity,
a special configuration of the system was set up by adding a linear Glan 
polariser ${\rm Glan}_{\rm ex}$ (Fig.~\ref{shema-general}) 
between the exit window of the cavity and the first mirror.
Such a device polarises the beam in a completely linear state ${\bf E}_G$ 
before it enters the optical elements of the exit line.
In this configuration, a number of data samples ${N_d}$ were recorded 
using the ellipsometer, each with a different azimuthal angle $\phi_G^{\ell}$ 
(${\ell=1,\cdots,N_d}$) of the polariser ${\rm Glan}_{\rm ex}$. 
For each $\phi_G^{\ell}$, the ellipsometer QWP was rotated azimuthally from 
$0^\circ$ to $360^\circ$.

A fit to these data samples is performed by minimising 
the $ \chi^2$ function defined in Eq.(\ref{eq:chi2pola}) to obtain
the only free parameters ${\xi_{\rm elli}^{\ell}}$ and 
${\phi_{\rm elli}^{\ell}}$ (${\ell=1,\cdots,N_d}$) of the beam polarisation
state after the HBS.
Using the resulting ${\xi^{\ell}_{\rm elli}}$ and ${\phi^{\ell}_{\rm elli}}$ 
and their uncertainties
${\sigma_{\xi^{\ell}_{\rm elli}}}$ and ${\sigma_{\phi^{\ell}_{\rm elli}}}$,
the following $ \chi^2$ function is constructed
in order to determine the elements of the matrix ${ M_T}$:
\begin{equation}\label{chi2biref}
{ \chi^2\ =\ \sum_{\ell =1,N_d}\ }\left[
\left( {\frac{\xi^{\ell}_{\rm th}-\xi^{\ell}_{\rm elli}}{\sigma_{\xi^{\ell}_{\rm elli}}}} \right)^2 +
\left( {\frac{\phi^{\ell}_{\rm th}-\phi^{\ell}_{\rm elli}}{\sigma_{\phi^{\ell}_{\rm elli}}}} \right)^2
\right]\,,
\end{equation}
where ${ \xi^{\ell}_{\rm th}}$ and ${\phi^{\ell}_{\rm th}}$ are 
the theoretical angles defining the ${\ell^{\rm th}}$ polarisation state 
after the HBS. By using Eqs.(\ref{eq:matricepgr}),(\ref{eq:descri-biref}),
these angles can be written as functions of the parameters ${\theta_1}$, 
${\theta_2}$, ${\theta_3}$, ${\gamma_1}$, ${\gamma_2}$, ${ p_1}$ and ${ p_2}$
of the matrix ${ M_T}$ and of the angle ${\phi_G^{\ell}}$ of the 
linear initial polarisation state.
A Monte Carlo study of the $\chi^2$ function defined in Eq.(\ref{chi2biref})
shows that some elements of the matrix ${ M_T}$ are completely
correlated and that the system can be described only with one delay plate,
one partial polariser and two rotators.
The parameters of ${ M_T}$ in Eq.(\ref{eq:descri-biref}) are
therefore restricted to ${\gamma_1 \equiv \gamma_T}$, ${p_1 \equiv p_T}$, 
${p_2=1}$, ${\theta_3=0}$ and ${\gamma_2 = 0}$.
The minimisation of the ${\chi^2}$ defined in Eq.(\ref{chi2biref}) leads to
$\gamma_T=  (13.1 \pm 1.4)\,{\rm mrad}$ and
$p_T = 1.001 \pm 0.001$, 
thereby showing that the exit optical system behaves like a delay plate 
inducing a birefringence of the order of a few tens of mrad.

A cross-check of this study has been performed 
by placing the additional polariser ${\rm Glan}_{\rm ex}$
between the HBS and the ellipsometer QWP.
As previously, ellipsometer
data sets were recorded for several azimuthal angles of the polariser.
Applying the same procedure, 
the matrix ${M_T}$ is now expected to be compatible with the identity matrix, since
there is no optical component between the Glan and the entrance of 
the ellipsometer.
The result of the fit, with an angle of
$(1.5 \pm 4.0)\,{\rm mrad}$ for the delay plate and a value of
${1.004 \pm 0.005}$ for the partial polariser parameter,
constitutes a valuable check of the robustness of our model describing
the exit beam line of the Fabry-Perot cavity.

\subsubsection{${S_3}$ at the exit of the cavity}\label{exit-cav}

Removing the polariser ${\rm Glan}_{\rm ex}$ to recover the standard setup 
of the optical system, the degree of circular polarisation ${S_3^{\rm ex}}$ 
at the exit of the cavity
has now to be determined as a function of ${S_3^{\rm elli}}$ 
at the entrance of the ellipsometer (Sect.~\ref{docp-measurements}).
The polarisation state $ {\bf E}_{\rm ex}$ at the exit of the cavity
is related to ${\bf E}_{\rm elli}$ (Eq.(\ref{eq:docp}))
as ${{\bf E}_{\rm ex}= M_T^{-1} \ {\bf E}_{\rm elli}}$.
The development of this field expression leads to the relation:
\begin{equation}\label{eq:biaisF}
{S_3^{ex}= S_3^{\rm elli}\ +\ 
\delta S_{\rm ex}(\theta_1,\theta_2,\gamma_T,p_T,\xi_{\rm elli},\phi_{\rm elli})}\,.
\end{equation}
The relation (\ref{eq:biaisF}) applied to several values of ${S_3^{\rm elli}}$
determined by the ellipsometer shows that the correction values
$\delta S_{\rm ex}$ are all below five per mill.
As the HBS is not the cause of this parasitic ellipticity
(see Sect.~\ref{exp-setup}), the bias
is due to the two mirrors system. 
This confirms a measurement performed at Saclay in 1999 for the TJNAF 
polarimeter~\cite{mirror-saclay} in which the
effect of the two mirrors on the determination of ${S_3}$
was measured to be of the order of a few per mill.

The bias $\delta S_{\rm ex}$ is calculable for each value of 
${S_3^{\rm elli}}$ and therefore does not enter as a systematic error 
but is explicitly determined to correct ${S_3^{\rm elli}}$.
The uncertainty ${\sigma_{S_3^{\rm ex}}}$ of ${S_3^{\rm ex}}$ 
due to the transfer from
${S_3^{\rm elli}}$ to ${S_3^{\rm ex}}$ are of order of a few ${10^{-4}}$ and
has been calculated
from the ${M_T}$ elements as follows: 
for four combinations 
(${\gamma_T + \sigma_{\gamma_T}}$, ${ p_T + \sigma_{p_T}}$),
(${ \gamma_T - \sigma_{\gamma_T}}$, ${ p_T + \sigma_{p_T}}$), 
(${ \gamma_T + \sigma_{\gamma_T}}$, ${ p_T - \sigma_{p_T}}$) and 
(${ \gamma_T - \sigma_{\gamma_T}}$, ${ p_T - \sigma_{p_T}}$), 
the corresponding ${S_3^{\rm ex}}$
are extracted and ${\sigma_{S_3^{\rm ex}}}$ 
is taken to be the maximum difference between these four values 
with the central value ${S_3^{\rm ex}}$ being calculated with parameters 
${\gamma_T}$ and ${p_T}$.

\subsection{Overall $S_3$ uncertainty}

Summarising all the studies and results described previously,
the ${S_3}$ value inside the cavity can be written by taking into account 
all the uncertainties of the optical system as:
\begin{equation}
S_3\ =\  S_3^{\rm elli} + \delta S_{\rm ex} \pm
\sigma_{S_3^{\rm elli}} \pm
\sigma_{S_3^{\rm ex}} \pm
\sigma_{\rm time} \pm  \sigma_{\rm trans}\,,
\end{equation}
where ${S_3^{\rm elli}}$ is the degree of circular polarisation measured 
using the ellipsometer, $\delta S_{\rm ex}$
is the ${S_3^{\rm elli}}$ dependent correction factor  
defined in Eq.(\ref{eq:biaisF}),
${\sigma_{S_3^{\rm elli}}}$, of the order of a few ${ 10^{-4}}$, is 
the uncertainty on the measurement of ${S_3^{\rm elli}}$ 
using the ellipsometer (Sect.~\ref{docp-measurements}), 
${\sigma_{S_3^{\rm ex}}}$, of the order of a few ${ 10^{-4}}$, is 
the uncertainty on the determination of the transfer 
matrix ${ M_T}$ (Sect.~\ref{exit-cav}),
${ \sigma_{\rm time} \approx 3 \times 10^{-3}\ }$ is the conservative
uncertainty associated to the duration of of data taking of 
an ellipsometer data sample
(Sect.~\ref{docp-measurements}),
and ${ \sigma_{\rm trans} < 3 \times 10^{-5}}$ is the uncertainty related to
the passage of the light through the exit cavity mirror 
(Sect.~\ref{parasitic}).
The last two uncertainties $\sigma_{\rm time}$ and $\sigma_{\rm trans}$ 
are common to all ${S_3}$ measurements,
all others vary for each measurement of ${S_3}$.

\section{Coherence of ${{\bf S_3}}$ along the whole optical system}\label{aller-retour}

Although the previous studies have provided values of ${S_3}$ at 
the electron-laser IP with an uncertainty around three per mill,
the idea is now to characterise also the entrance optical elements by 
a matrix ${M_E}$, determine the values of ${S_3}$
at different places of the optical system to check their coherence
and make sure that no additional unknown large effect could induce a bias 
on ${S_3}$ at the center of the cavity.

\subsection{Determination of $M_E$}\label{me-determination}

The entrance beam line is described with the matrix ${M_E}$
(see Fig.~\ref{shema-general}) and is composed of a glass plate, two lenses
and four alignment mirrors.
Following the optical theorem of R.\ Clark Jones~\cite{jones-II-1941} 
already used in Sect.~\ref{exit-matrice},
${M_E}$ can be expressed with the same
formula (see Eq.(\ref{eq:descri-biref})) as for the matrix ${M_T}$.
To determine ${M_E}$, the method pursued is to model the passage of the beam 
from the entrance Glan polariser to the entrance cavity mirror when 
the cavity is unlocked,
followed by the retro-reflection of the beam by the cavity mirror and
its passage through the Glan in the opposite direction. 
A reversibility theorem~\cite{jones-I-1941,aspect-1993} states that 
for a matrix $ M $ describing the light path through a given system,
the matrix corresponding to the light path in the opposite direction is 
the transposed matrix of $ M $.
Following this theorem and starting with a horizontal linear polarisation 
state  ${\bf E}_{\rm lin} = ( 1, 0)^T$ just after the entrance Glan,
the expression of the retro-reflected field ${\bf E}_{\rm ret}$ and 
the associated intensity ${T_{\rm ret}}$ emerging from the Glan in the
return direction can then be written as:
\begin{equation}\label{eq:entrance-ar}
\begin{array}{ll}
& {\bf E}_{\rm ret}= G_v M_{\rm QM}^T M_m M_{\rm QM} {\bf E}_{\rm lin}\,,\hspace{5mm}
T_{\rm ret}= \left| {\bf E}_{\rm ret} \right|^2 \\
{\rm with} \hspace{10mm} & M_{\rm QM} = M_E R_{\rm QE} M_{\rm QW} R_{\phi_{\rm ent}}\,,
\end{array} 
\end{equation}
where ${G_v}$ is the matrix of the Glan polariser allowing only the vertical 
component of the field to pass when the beam returns,
${M_{\rm QM}}$ is the transfer matrix of the optical line
from the plate ${\rm QWP}_{\rm ent}$ to the last alignment mirror,
${ M_m}$ is the Jones matrix of the entrance cavity mirror,
${M_E}$ is the transfer matrix to be determined,
${M_{\rm QW}}$ is the Jones matrix of the plate ${\rm QWP}_{\rm ent}$, and
${R_{\rm QE}}$ and ${R_{\phi_{\rm ent}}}$ are two 
${2 \times 2}$ rotation matrices introduced to reflect
the azimuthal orientation of ${\rm QWP}_{\rm ent}$ with respect to 
the matrix ${M_E}$ and to the Glan polariser axes, respectively.

According to this modelisation, the elements of ${M_E}$ have been determined
from data recorded with the photodiode ${\rm pd}_{\rm ent}$ 
for various positions ${\phi_{\rm ent}}$ of the plate ${\rm QWP}_{\rm ent}$, 
by minimising the following $\chi^2$ function:
\begin{equation}\label{eq:chi2-entr}
\chi^2\ =\ \sum_{i=1,N_{\rm ent}} \left( 
\frac{R\ T_{\rm ret}^i - I_{\rm ret}^i}
{\sigma_{I_{\rm ret}^i}} \right)^2\,,
\end{equation}
where ${N_{\rm ent}}$ is the number of different azimuthal positions 
${\phi_{\rm ent}}$, ${T_{\rm ret}^i}$ (${I^i_{\rm ret}}$) is the theoretical 
(experimental) intensity calculated with Eq.(\ref{eq:entrance-ar}) 
(measured with ${\rm pd}_{\rm ent}$)
at the ${i^{\rm th}}$ value of ${\phi_{\rm ent}}$,
${\sigma_{I_{\rm ret}^i}}$ is the uncertainty of ${I^i_{\rm ret}}$,
and $R$ is a normalisation factor which is determined
by solving ${\partial \chi^2 / \partial R =0}$.
When the system was conceived, the characterisation of the entrance optical 
line was not planned and
the photodiode ${\rm pd}_{\rm ent}$ was only devoted to find the 
azimuthal positions of the plate ${\rm QWP}_{\rm ent}$
leading to a right or left circular polarisation of the laser beam.
No specific study has thus been conducted to
study the response and measurement uncertainties of this photodiode,
and in parti-
cular no photodiode thermal regulation and no additional 
reference photodiode 
to compensate the laser power variations have been installed.
A measurement ${I^i_{\rm ret}}$ using the photodiode ${\rm pd}_{\rm ent}$ 
thus corresponds simply to
the mean value over ten thousand signals recorded with a $12$-bit 
analog-to-digital converter at a sample rate of $2\,{\rm MHz}$.
The uncertainty ${\sigma_{I^i_{\rm ret}}}$ 
is defined as the RMS value of the distribution of ${I^i_{\rm ret}}$ 
and is of the order of one to two percent.
This level of precision is not as good as the one obtained with the
ellipsometer photodiodes
as described in Sect.~\ref{elli-component}, 
and consequently, the development of a complete theoretical model to describe 
each optical component would not make
sense here. It is therefore sufficient to consider the Glan polariser,
the plate ${\rm QWP}_{\rm ent}$ (which is a quartz plate treated with 
an anti-reflection coating)
and the cavity mirror as perfect and to write the corresponding expression of 
the Jones matrices used in Eq.(\ref{eq:entrance-ar}) as:
\begin{equation}\label{matrix-perfect}
{G_v = }\left( \begin{array}{cc} 0 & 0 \\ 0 & 1 \end{array} \right)\,,\hspace{5mm}
{M_m =} \left(\begin{array}{cc} 1&0 \\ 0&0 \end{array}\right)\,,\hspace{5mm}
{M_{\rm QW} = }\left( \begin{array}{cc}
 1 & 0 \\
 0   & {e^{-{\it i}\pi/2}}
\end{array} \right)\,.
\end{equation}

\begin{figure}[htbp]
\begin{center}
\includegraphics[width=0.55\textwidth]{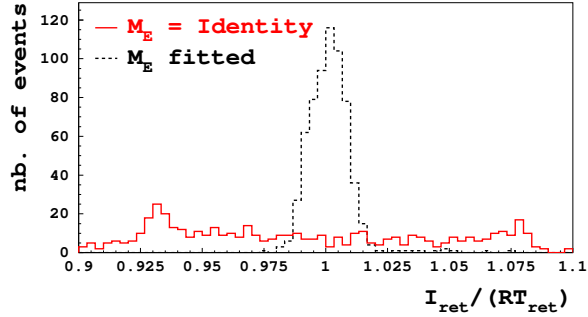}
\end{center}
\caption{Ratio of the measured intensities 
over the theoretical ones ${I_{\rm ret}/(RT_{\rm ret})}$ comparing the case
where the $M_E$ is from the fit (dotted line) and the case where the ${M_E}$ 
is fixed to the identity (full line).}
\label{i-entrance}
\end{figure}

A Monte Carlo study shows that our data are well described by using only 
one delay plate and one polariser. 
The parameters in Eq.(\ref{eq:descri-biref}) are
therefore restricted to ${\gamma_1 \equiv \gamma_E}$, ${p_1 \equiv p_E}$, 
${p_2=1}$, ${\theta_3=0}$ and ${\gamma_2 = 0}$, and
the minimisation of the $\chi^2$ defined in Eq.(\ref{eq:chi2-entr}) leads to 
$\gamma_E =  (-32.8 \pm 0.5)\,{\rm mrad}$ and ${p_E = 1.17 \pm 0.01}$.
The effect of the matrix ${M_E}$ is clearly visible in Fig.~\ref{i-entrance} 
showing the distribution of the ratio ${I_{\rm ret}/(RT_{\rm ret})}$ 
either for the case where the minimisation is performed (dotted line) or 
for the case where the 
matrix ${M_E}$ is fixed to the identity (full line). 

\subsection{Coherence of {\bf $S_3$}}

Using the matrices ${M_T}$ and ${M_E}$,
the ${S_3}$ values at the four different locations
${S_3^{\rm ent}}$, ${S_3^{\rm in}}$, ${S_3^{\rm ex}}$ and ${S_3^{\rm elli}}$
indicated in Fig.~\ref{shema-general} can now be determined for 
any polarisation state of the laser
beam, {\it i.e.} for any azimuthal position ${\phi_{\rm ent}}$ of the 
motorised rotating mount ${\rm QWP}_{\rm ent}$.
These four values are determined from their associated electric fields
${{\bf E}_{\rm ent}}$, ${{\bf E}_{\rm in}}$, ${{\bf E}_{\rm ex}}$ and 
${{\bf E}_{\rm elli}}$.
We recall that ${{\bf E}_{\rm ent}}$ is calculated from a linearly polarised 
beam passing through the plate ${\rm QWP}_{\rm ent}$ which is positioned at 
the azimuthal angle ${\phi_{\rm ent}}$ with respect to the
entrance Glan axis:
${ {\bf E}_{\rm ent}= M_{QW}\ R_{\phi_{ent}}\ (1,0)^T}$,
${{\bf E}_{\rm in}}$ is derived from ${{\bf E}_{\rm ent}}$ and 
from the matrix ${M_E}$:
${ {\bf E}_{\rm in}= M_E\ {\bf E}_{\rm ent}}$,
${{\bf E}_{\rm elli}}$
is determined as described in Sect.~\ref{docp-measurements}
from an ellipsometer data sample, and
${{\bf E}_{\rm ex}}$ is calculated from the field ${{\bf E}_{\rm elli}}$
and the matrix ${M_T}$ as:
${{\bf E}_{\rm ex}= M_T^{-1}\ {\bf E}_{\rm elli}}$.

\begin{figure}[t]
\begin{center}
\includegraphics[width=0.60\textwidth]{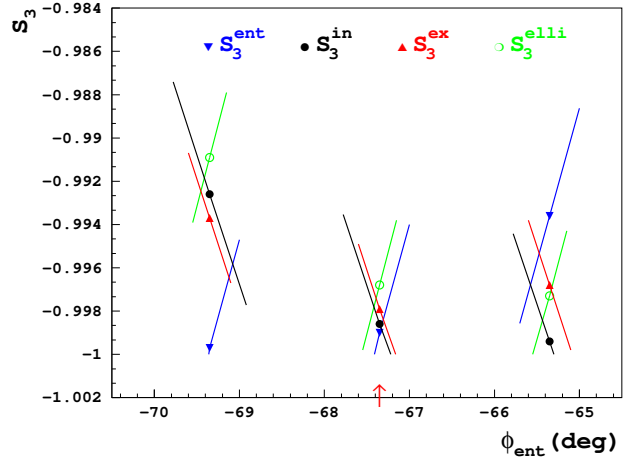}
\end{center}
\caption{${S_3^{\rm ent}}$ (blue triangles), ${S_3^{\rm in}}$ (black points), 
${S_3^{\rm ex}}$ (red triangles), and ${S_3^{\rm elli}}$ (open green circles) 
for three azimuthal positions of the plate ${\rm QWP}_{\rm ent}$ around a left
circularly polarised state
(indicated by an arrow). The error bars are inclined for clarity.}
\label{docp-vs-phi}
\end{figure}

To check experimentally the coherence of ${S_3}$ along the optical system,
several arbitrarily va-
lues of ${\phi_{\rm ent}}$
have been chosen close to
a circularly polarised light state and, for each one of these positions, 
a data sample has been recorded in the ellipsometer as described in 
Sect.~\ref{docp-measurements}.
The evolution of ${S_3}$ along the optical path can be followed in 
Fig.~\ref{docp-vs-phi}
through the values of ${S_3^{\rm ent}}$, ${S_3^{\rm in}}$, ${S_3^{\rm ex}}$
and ${S_3^{\rm elli}}$ presented for three positions ${\phi_{\rm ent}}$
around a left circularly polarised state.
In Fig.~\ref{docp-vs-phi}, an uncertainty of $0.5\%$ on the values of ${S_3^{\rm ent}}$
is taken (typical known value as mentioned in the introduction).
This uncertainty propagates directly to that of ${S_3^{\rm in}}$.

For the measurement of the lepton beam polarisation, 
the only relevant quantity is the light polarisation
inside the cavity, to which, of course, we do not have access but 
which is located
between the two values ${S_3^{\rm in}}$ and ${S_3^{\rm ex}}$.
As shown in Fig.~\ref{docp-vs-phi}, 
the difference $|S_3^{\rm ex}-S_3^{\rm in}|$
is less than one per mill when ${S_3^{\rm ex}}$ is closer to ${- 1}$
({\it i.e.}\ when the system is at its operating point~\cite{polca-paper1}),
and can reach up to three per mill
in the explored domain of ${\phi_{\rm ent}}$.
Part of the difference could be explained by the presence of 
a small birefringence
due to multi-layers coating cavity mirrors as mentioned
in Sect.~\ref{parasitic}.
We do not know the exact value of our mirror coating birefringence, but
birefringences have been measured for instance 
in~\cite{biref-mirror-coating-jacob-et-al-1995,biref-mirror-coating-lee-et-al-2000,biref-mirror-coating-brandi-et-al-1997} 
for cavity finesses of
$6\,600-100\,000$.
In all these measurements the order of magnitude
of the birefringence is a few ${10^{-6}}\,{\rm rad}$. 
The Fabry-Perot cavity, with its multi-layer coating mirrors, has
a finesse of about $30\,000$~\cite{polca-paper1} and thus lies 
within the range quoted above.
Because of the resonant optical cavity, the phase shift due to a single 
passage of the light
in the reflected coating is amplified by a factor 
${2 {\mathcal F} /\pi}$~\cite{biref-mirror-coating-brandi-et-al-1997}
and becomes of the order of a few $10^{-2}\,{\rm rad}$. 
The bias on ${S_3}$ can be expressed in term of this amplified
birefringence ${\phi_{\rm bir}}$ as 
${S_3^{\rm ex}-S_3^{\rm in} \approx \phi_{bir}^2/2}$~\cite{habil-marie} and 
can therefore be of a few per mill.
However, another systematic source, which could explain the
difference of a few per mill between ${S_3^{\rm in}}$ and ${S_3^{\rm ex}}$, 
is the lack of precision in measurements with ${\rm pd}_{\rm ent}$ used 
for the determination of ${M_E}$ (Sect.~\ref{me-determination}) and 
thus of ${S_3^{\rm in}}$.
Anyway, the study of the entrance beam line does not intend to give 
an accurate measurement of ${S_3^{\rm in}}$ but is devoted to check 
the coherence of the system and particularly the coherence of
measurements just before and after the cavity.

\section{Summary}

The implementation of an uncoated QWP in the ellipsometer
of the Fabry-Perot cavity polarimeter of HERA has allowed us to determine 
the degree of circular
polarisation ${S_3}$ of the laser at the entrance of the ellipsometer
with an uncertainty of $0.3\%$. Such a small uncertainty is achieved
thanks to a complete model description
of the ellipsometer optical system.
The transport of ${S_3}$ up to the electron-laser IP has then been 
studied and the modeling of the optical elements located between the IP 
and the ellipsometer has made it possible 
to conserve the uncertainty of $0.3\%$ at the IP.
A study of the optical line
from the IP up to the laser head has also been performed
and has shown that even with an unoptimised photometric measurement, 
${S_3}$ is controlled along the optical path 
at the few per mill level.
The level of accuracy presented here has, to our knowledge, never been reached
in the environment of a particle collider and
provides a good prospect for applications in a future linear 
collider~\cite{three-polarimeter-et-tesla-2002,ref-schuler,gudi}.

\section*{Acknowledgment}
We would like to thank T.~Cac\'er\`es, M.~Delbard, N.~Falletto, 
M.~Linz, A.~Reboux and M. Woods for their help to this work.

\end{document}